# Subject Enveloped Deep Sample Fuzzy Ensemble Learning Algorithm of Parkinson's Speech Data

Yiwen Wang #, Fan Li #, Xiaoheng Zhang#, Pin Wang, Yongming Li *

(School of Microelectronics and Communication Engineering, Chongqing University, Chongqing 400044, P.R. China)

# as co-first author, * as corresponding author

**Abstract**
Parkinson disease (PD)'s speech recognition is an effective way for its diagnosis, which has become a hot and difficult research area in recent years. As we know, there are large corpuses (segments) within one subject. However, too large segments will increase the complexity of the classification model. Besides, the clinicians interested in finding diagnostic speech markers that reflect the pathology of the whole subject. Since the optimal relevant features of each speech sample segment are different, it is difficult to find the uniform diagnostic speech markers. Therefore, it is necessary to reconstruct the existing large segments within one subject into few segments even one segment within one subject, which can facilitate the extraction of relevant speech features to characterize diagnostic markers for the whole subject. To address this problem, an enveloped deep speech sample learning algorithm for Parkinson's subjects based on multilayer fuzzy c-mean (MlFCM) clustering and interlayer consistency preservation is proposed in this paper. The algorithm can be used to achieve intra-subject sample reconstruction for Parkinson's disease (PD) to obtain a small number of high-quality prototype sample segments. At the end of the paper, several representative PD speech datasets are selected and compared with the state-of-the-art related methods, respectively. The experimental results show that the proposed algorithm has higher recognition accuracy and better generalization performance. The proposed method in this paper can effectively reconstruct all speech sample segments within Parkinson's subjects and obtain a small number of prototype samples reflecting the overall pathological status of the subjects. This not only facilitates the extraction of speech features (diagnostic markers) reflecting the overall pathological status of the subjects, but also significantly improves the classification accuracy and generalization performance, and better meets the requirements of clinical applications.

1. INTRODUCTION

Parkinson's disease is a common neurodegenerative disease. The existing treatments can only alleviate the symptoms and cannot stop the progression of the disease [1], which makes early PD diagnosis particularly important. Most of the current PD diseases are diagnosed using functional imaging and molecular biology, but the cost of the testing instruments and some of the tests required for these diagnostic modalities are high and cannot yet be routinely performed. Therefore, there is a real-life need for a low-cost, efficient and universally applicable way of PD diagnosis [2-3]. Speech recognition-based PD diagnosis methods have received a lot of attention because of their non-invasive and high cost efficiency. In recent years, a large number of scholars at home and abroad have conducted research on speech recognition PD diagnosis and made some progress [4-8].

PD speech recognition methods are well researched, but they mainly involve feature learning and classifier design aspects. In terms of feature learning, there are two types of feature selection and feature extraction. Feature selection mainly contains three types: filtered, wrapped, and embedded. Its principle is mainly to select some most effective features from the original features as a new feature subset, so as to reduce the dataset dimension and improve the learning algorithm performance. Feature extraction is different from feature selection, which is mainly through data transformation or data mapping to get a new feature space. The meaning of the data in the new feature space is different from the original data, but contains most of the features of the previous data, and has a lower dimensionality, which is convenient for further analysis. The main methods are PCA (principle component analysis) [9-11], LDA (linear discriminant analysis) [12-14], and LPP (locality preserving projection) [15], neural network (NN) [16,17]. In terms of classifier design, it is desired that the classifiers have low complexity and high generalization ability. The main classifiers contain SVM (support vector machine) [18,19], KNN (K-nearest neighbor) [20,21], ELM (extreme learning machine) [22,23], RF (random forests) [24,25], etc.

At present, research on PD speech diagnosis still faces the following clinical problems and needs (challenge).

1) It is difficult to reflect the Parkinson's pathological state comprehensively with a single corpus, and ensure the accurate diagnosis of PD speech since people have no enough priori knowledge. A large number of corpus samples (sample segments) need to be collected for each subject during PD speech diagnosis. However, when the number of samples collected for each subject is large, it will lead to a high complexity of the classification model and poor generalization ability. Therefore, it is necessary for sample reconstruction to reduce the number of segments collected for each subject, and improve the generalization ability of the classification model and the efficiency of PD speech diagnosis.

2) In the actual PD speech diagnosis, clinicians are interested in finding diagnostic speech markers that reflect the pathology of the whole subject. However, there are various segments within one subject, and the optimal relevant features of each speech sample segment are different to some extent. It is difficult for clinicians to find uniform diagnostic markers that reflect the pathology of the subjects. Therefore, it is necessary to combine the large segments into few segments, thereby finding the uniform optimal features (diagnostic markers).

These studies are valuable but do not provide a good answer to the clinical problems and needs mentioned above. Several scholars have attempted to solve the problems. Sakar et al. [18]. performed a linear transformation on the speech sample segments of subjects, using centralized trend and discrete measures such as mean, median, trimmed mean (removing 10% and 25%), standard deviation, interquartile range, and mean absolute deviation, for 26 speech samples from each subject, resulting in a small number of sample segments to improve the generalization ability and classification accuracy of the model. However, the transformation is simple and has limitation on sample reconstruction.

The possible major reasons are as: 1) the linear transformation is simple for sample transformation compared with nonlinear transformation. 2) the method is based on one layer of transformation. Therefore, it is necessary to design one method for multiple layer of nonlinear transformation of the PD speech segments, thereby constructing small number of high quality of new sample segments. Deep learning method inspires us. The kind of method can obtain new features of higher quality by performing multilayer nonlinear transformation on the original features [26,27]. Inspired by this method, we propose a deep sample learning method based on multilayer fuzzy c-mean (FCM) clustering [28] to reflect the hierarchical structure information between samples. To avoid the problem of the inconsistent distribution of the generated new samples and the original samples, the minimum interlayer discrepancy mechanism based on maximum mean discrepancy (MIDMD) is introduced on the basis of FCM to make the deep samples consistent with the original sample distribution [29,30]. To further obtain a small number of high-quality prototype samples, we perform sample pruning within subjects. Besides, simultaneous selection of sample segments and features mechanism is considered here.

The proposed algorithm can be used to achieve intra-subject sample reconstruction for Parkinson's disease (PD), resulting in a small number of high-quality prototype samples and solving the problems mentioned above. The main contributions and innovations of this paper are as follows.

1) An enveloped deep sample learning network is proposed based on MlFCM and MIDMD-MlFC&IDMD: The MlFCM model is designed, and the interlayer consistency maintenance mechanism is constructed through the MIDMD. Combining the two parts above, an enveloped deep sample learning network is designed, thus obtaining a deep sample space characterizing the original samples, and has obtained a small number of high quality new samples.

2) The ensemble algorithm of PD speech data is designed based on MlFC&IDMD and SP module (called MlFC & SP_IDMD) is proposed: The sample set was pruned and then constructed as an envelope with all sample segments of a single subject. The envelope was processed using the deep sample learning network MlFC & IDMD to construct a PD speech sample segment deep space. The Sample segments and features in the envelop after MIFC&IDMD are simultaneously selected and sub-classifiers are constructed based on the selected segments and features. Finally, decision layer fusion is performed on all the sub-classifiers.

3) Based on the algorithm proposed in this paper, not only the classification accuracy and generalization performance can be improved, but also the features (diagnostic markers) reflecting the overall pathological status of the subject can be extracted. This helps physicians understand the mechanism of identifying PD by speech and also helps improve clinical applications. In other words, this proposed algorithm solves the clinical problems and needs mentioned above to some extent.

## 2. PROPOSED METHOD

### 2.1 Notation

| Symbol | Description |
|---|---|
| $S$ | Original dataset |
| $\overline{S}$ | Based on the dataset obtained after subject transposition |
| $\hat{S}_p$ | Dataset after sample pruning |
| $P$ | The new set of samples obtained after clustering |
| $P^L$ | The dataset of the Lth level of the depth sample space |
| $U$ | Affiliation Matrix |
| $T$ | Dataset based on subjects' samples after stitching |
| $E$ | Predictive label matrix for deep sample space |
| $NH(s_i)$ | $s_i$ similar nearest neighbor vectors |
| $NM(s_i)$ | $s_i$ dissimilar nearest neighbor vector |
| $W$ | Weight vector |
| $N$ | Number of samples in the original dataset |
| $\overline{N}$ | Number of samples in the dataset after pruning |
| $L$ | Number of depth sample space layers |
| $C_L$ | Number of samples in the sample space of the Lth layer |
| $\beta$ | The weight vector of $E$ |

| $\beta'$ | Normalized $\beta$ |
|---|---|
| $y$ | The prediction results of the proposed algorithm |

## 2.2 Algorithm analysis (MlFC & SP_IDMD)

The main process of the proposed algorithm is described as follows: first, all the sample segments within one subject are put into one envelope for subsequent processing. Second, sample pruning transformation is performed based on the envelope to retain the sample segments that can characterize the pathology of the subject with high quality, and the envelope is updated after the sample pruning. Next, the deep sample learning mechanism designed based on multilayer FCM with MIDMD to obtain a deep sample space. After FCM and MIDMD, the envelope is renewed again. Based on the updated envelope of each layer of the deep sample space, the sample segments and features are simultaneously filtered to obtain a small set of high-quality samples and features (filtered envelope). Based on the filtered envelopes of every layer of the deep sample space, sub-classifiers are trained separately. Finally, all the results from the sub-classifiers are fused by sparse fusion mechanism.

### 2.2.1 Sample pruning module -SP Module

Suppose a dataset has $n$ subjects, $S = \begin{bmatrix} S_1 \\ S_2 \\ ... \\ S_n \end{bmatrix} \in R^{(n \times m) \times d}$ each subject has $m$ d-dimensional corpus samples

$S_l = \begin{bmatrix} s_{11} & s_{12} & ... & s_{1d} \\ s_{21} & s_{22} & ... & s_{2d} \\ ... & ... & ... & ... \\ s_{m1} & s_{m2} & ... & s_{md} \end{bmatrix} \in R^{m \times d}, l = (1, 2, ..., n)$, based on each subject transposed to obtain

$S_l^T = \begin{bmatrix} s_{11} & s_{21} & ... & s_{m1} \\ s_{12} & s_{22} & ... & s_{m2} \\ ... & ... & ... & ... \\ s_{1d} & s_{2d} & ... & s_{md} \end{bmatrix} \in R^{d \times m}$. Each feature vector $s_i = \begin{bmatrix} s_{1i} & s_{2i} & ... & s_{mi} \end{bmatrix}, i \in (1, 2, ..., n \times d)$ is selected in turn from the dataset, and the two nearest neighbors of the feature vector are traversed in the dataset: one is from the same class as $s_i$ as the nearest neighbor $NH(s_i)$, and the other is from a different class than $s_i$ as the nearest neighbor $NM(s_i)$. The Euclidean distance of this feature vector to the two nearest neighbors is calculated, and the $j \in (1, 2, ...m)$ dimension of the weight vector is updated iteratively by the following equation:

$$w_j = w_j + |s_{ij} - NM(s_{ij})| - |s_{ij} - NH(s_{ij})| \quad (1)$$

Where $|\cdot|$ is the absolute value notation, $w_j$ denotes the weight of the $j$ sample, $s_{ij}$ denotes the $j$ dimension of the feature, and $NH(s_{ij})$ and $NM(s_{ij})$ denote the $j$ dimension of the similar nearest neighbor and dissimilar nearest neighbor of the feature vector $s_i$, respectively.

The m-dimensional sample weight vector $W = \begin{bmatrix} w_1 & w_2 & ... & w_m \end{bmatrix} \in R^m$ is arranged in ascending order, and the sample segments corresponding to the lower weight values are subtracted according to the set pruning number *cutoff*. The pruned dataset is transposed by subjects to obtain the transformed dataset

$$\hat{S}_p = \begin{bmatrix} S_{p1} \\ S_{p2} \\ ... \\ S_{pn} \end{bmatrix} \in R^{(n \times (m-cutoff)) \times d} \quad .$$

### 2.2.2 Deep sample learning mechanism - MlFC&IDMD

The deep sample space is established based on the envelope formed by the subjects using the MIFCM algorithm, and the MIDMD is used as a constraint term to make the new dataset $P = \begin{bmatrix} p_{C1} \\ p_{C2} \\ ... \\ p_{Cn} \end{bmatrix} \in R^{(n \times C) \times d}$ in the sample space layer consistent with the distribution of the envelope formed by the sample segments of the original dataset $\hat{S}_p = \begin{bmatrix} S_{p1} \\ S_{p2} \\ ... \\ S_{pn} \end{bmatrix}$ corresponding to the subjects, so that the new objective function obtained is:

$$\min F_1(U, P) = \sum_{i=1}^{C} (u_{ik})^m (d_{ik})^2 + \lambda \left( \sum_{i=1}^{C} u_{ik} - 1 \right) + L_{\Psi_H}\left( \hat{S}_p, P \right) \tag{2}$$

The above equation can be written as follows.

$$\min F_1(U, P) = \min \left( F(U, P) + F_{\Psi_H}(U, P) \right) \tag{3}$$

Among them:

$$F(U, P) = \sum_{i=1}^{C} \sum_{k=1}^{N} (u_{ik})^m (d_{ik})^2 + \lambda \left( \sum_{i=1}^{C} u_{ik} - 1 \right) \tag{4}$$

$$F_{\Psi_H}(U, P) = \left\| \frac{1}{N^2} \sum_{i=1}^{N} \sum_{i'=1}^{N} k\left( s_{pi}, s_{pi'} \right) - \frac{2}{NM} \sum_{i=1}^{N} \sum_{j=1}^{M} k\left( s_{pi}, p_j \right) + \frac{1}{M^2} \sum_{j=1}^{M} \sum_{j'=1}^{M} k\left( p_j, p_{j'} \right) \right\| \tag{5}$$

Let $K_{(s_{pi}, s_{pi'})} = \sum_{i=1}^{N} \sum_{i'=1}^{N} k_{(s_{pi}, s_{pi'})}$, then simplify Eq. (5) to Eq. (6).

$$F_{\Psi_H}(U, P) = \left\| \frac{1}{N^2} K_{(s_{pi}, s_{pi'})} - \frac{2}{NM} K_{(s_{pi}, p_j)} + \frac{1}{M^2} K_{(p_j, p_{j'})} \right\| \tag{6}$$

Let $K' = \begin{bmatrix} K_{(s_{pi}, s_{pi'})} & K_{(s_{pi}, p_j)} \\ K_{(s_{pi}, p_j)} & K_{(p_j, p_{j'})} \end{bmatrix}$, $L = \begin{bmatrix} \dfrac{1}{N^2} & -\dfrac{1}{NM} \\ -\dfrac{1}{NM} & \dfrac{1}{M^2} \end{bmatrix}$, then simplify Eq. (6) to Eq. (7).

$$F_{\Psi_H}(U,P) = tr(K'L') \tag{7}$$

Solve for the minimalist solution of Eq. (5).

$$\begin{cases} \dfrac{\partial F_1(U,P)}{\partial u} = \dfrac{\partial F(U,P)}{\partial u} + \dfrac{\partial F_{\Psi_H}(U,P)}{\partial u} = \dfrac{\partial F(U,P)}{\partial u} = 0 \\ \dfrac{\partial F_1(U,P)}{\partial p} = \dfrac{\partial F(U,P)}{\partial p} + \dfrac{\partial F_{\Psi_H}(U,P)}{\partial p} = 0 \\ \dfrac{\partial F_1(U,P)}{\partial \lambda} = \left(\sum_{i=1}^{C} u_{ik} - 1\right) = 0 \end{cases} \tag{8}$$

The Eq. (8) can be solved by:

$$\frac{\partial F_1(U,P)}{\partial p_j} = -2\sum_{k=1}^{N}(u_{jk})^m (s_{pk} - p_j) - \frac{2}{NC}\sum_{k=1}^{N} s_{pk} + \frac{2}{C^2}\sum_{j=1}^{C} p_j \tag{9}$$

The sample transformation relationship between layers is obtained from Eq. (9) and the transformation relationship between the new sample $P$ and the original sample $x$.

$$P = A^{-1}\begin{bmatrix} b_1 \\ b_2 \\ \dots \\ b_C \end{bmatrix} \tag{10}$$

Among them:

$$A = \begin{bmatrix} \sum_{k=1}^{N}(u_{1k})^m + \dfrac{1}{C^2} & \dfrac{1}{C^2} & \dots & \dfrac{1}{C^2} \\ \dfrac{1}{C^2} & \sum_{k=1}^{N}(u_{2k})^m + \dfrac{1}{C^2} & \dots & \dfrac{1}{C^2} \\ \dots & \dots & \dots & \dots \\ \dfrac{1}{C^2} & \dfrac{1}{C^2} & \dots & \sum_{k=1}^{N}(u_{Ck})^m + \dfrac{1}{C^2} \end{bmatrix} \tag{11}$$

$$b_i = \sum_{k=1}^{N}\left[(u_{ik})^m + \frac{1}{NC}\right] * s_{pk}, \quad i = 1,2,\dots,C \tag{12}$$

Thus, the MlFC&IDMD algorithm is used to construct the L-layer deep sample space that maintains the consistency of the distribution between layers.

### 2.2.3 Simultaneous sample/feature selection mechanism - SS/FSM

In order to ensure the physical significance of the features and facilitate the mining of the synergistic relationship between sample segments and features, the algorithm in this paper considers the simultaneous sample/feature selection mechanism instead of feature extraction. The sample/feature simultaneous selection mechanism is described as follows.

Multiple sample segments are stitched together based on each subject $S_l = \begin{bmatrix} s_1 \\ s_2 \\ \dots \\ s_m \end{bmatrix} \in R^{m \times d}, l = (1,2,\dots,n)$ to obtain a new high-dimensional sample $S_l = \begin{bmatrix} s_1 & s_2 & \ldots & s_m \end{bmatrix} \in R^{1 \times (m \times d)}$. The features of this new sample consist of the features of all sample segments of this one subject, and the dataset composed of these new high-dimensional samples is denoted $T = \begin{bmatrix} S_1 \\ S_2 \\ \ldots \\ S_n \end{bmatrix} = \begin{bmatrix} t_{11} & t_{12} & \ldots & t_{1(m \times d)} \\ t_{21} & t_{22} & \ldots & t_{2(m \times d)} \\ \ldots & \ldots & \ldots & \ldots \\ t_{n1} & t_{n2} & \ldots & t_{n(m \times d)} \end{bmatrix} \in R^{n \times (m \times d)}$. Let $T_l = \begin{bmatrix} t_{l1} & t_{l2} & \ldots & t_{l(m \times d)} \end{bmatrix}$. Select a sample $T_l$ from this dataset in turn, and iteratively find a similar nearest neighbor $NH(T_l)$, a dissimilar nearest neighbor $NM(T_l)$ of this sample in this dataset, and update the high-dimensional feature weights by the above Eq. (1). The weight vectors are sorted in ascending order and the sample features with the lowest weights are subtracted in turn.

*2.2.4  Sparse fusion mechanism*

Based on each layer of the deep sample space, the samples are divided into a training set and a test set. The training set is used to train the classifier of each layer, and the test set is used to test the prediction labels of each layer. The matrix composed of the prediction labels is denoted as $E = \begin{bmatrix} e_{11} & e_{12} & \ldots & e_{1L} \\ e_{21} & e_{22} & \ldots & e_{2L} \\ \ldots & \ldots & \ldots & \ldots \\ e_{n1} & e_{n2} & \ldots & e_{nL} \end{bmatrix} \in R^{n \times L}$, $L$ is the number of fusion layers. The objective function of this decision layer fusion mechanism is:

$$\min_{\beta}\left(\|y\text{-}E\beta\|_2^2 + \lambda \|\beta\|_1\right) = \min_{\beta}\left[\sum_{i=1}^{n}\left(y_i - \sum_{j=1}^{L} e_{ij}\beta_j\right)^2 + \lambda \sum_{j=1}^{L} |\beta_j|\right] \quad (13)$$

In the above equation, $y = \begin{bmatrix} y_1 \\ y_2 \\ \ldots \\ y_n \end{bmatrix} \in R^{n \times 1}$ is the true label, $\beta = \begin{bmatrix} \beta_1 \\ \beta_2 \\ \ldots \\ \beta_L \end{bmatrix} \in R^{L \times 1}$ is the weight vector, and $\lambda$ is the penalty coefficient for parameter estimation. The weight of each layer label is calculated by minimizing the objective function model, and the weight vector is normalized to obtain the vector $\beta' = \begin{bmatrix} \beta_1' \\ \beta_2' \\ \ldots \\ \beta_L' \end{bmatrix} \in R^{L \times 1}$. Based on this weight vector, the final predicted labels are obtained by weighting and summing the labels of each layer.

$$y = \Phi(E\beta') \in R^{n \times 1} \quad (14)$$

In the above equation, $\Phi(a) = \begin{cases} 0, a \leq 0.5 \\ 1, a > 0.5 \end{cases}$.

*2.3  The proposed algorithm*

The flowchart of the algorithm is described as follows. As described in the first part of section 2.2, the proposed algorithm is mainly made up of SP module, MIFCM&IDM, and SS/FSM. All the sample segments within one subject are put into one envelope. The envelope is updated during several stages by the SP module, MIFCM&IDM, and SS/FSM, and is concentrated. The proposed algorithm is helpful for improving classification accuracy, generalization capability and extracted features for the whole subject.

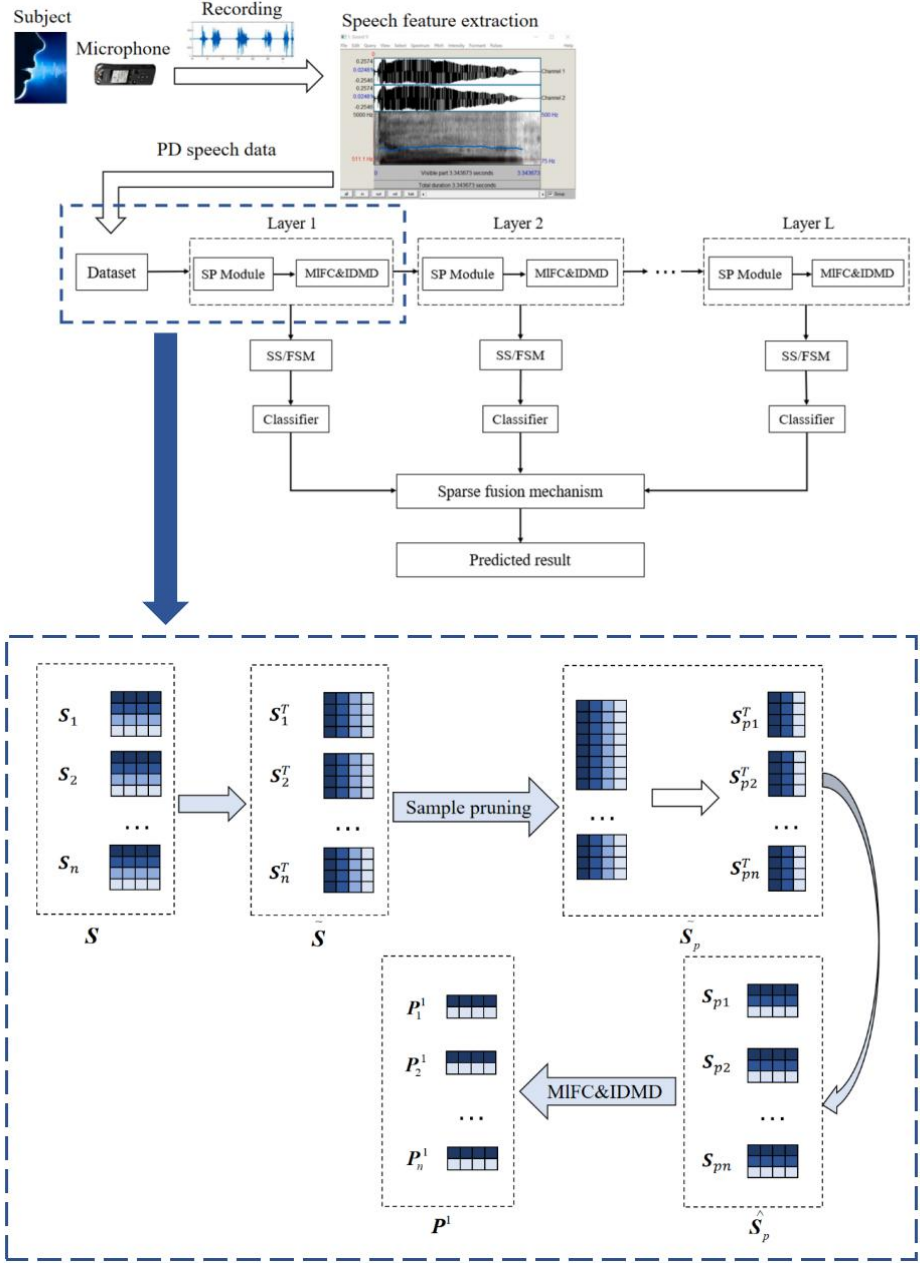

Fig.1. The flowchart of the proposed algorithm

The pseudo code is described as follows.

***Algorithm 1: MlFC & SP_IDMD***

**Input:** Original dataset $S$, Number of samples in the sample space of the i-th layer deep $C_i$, number of features per sample $m$, number of subjects n

**Output:** Accuracy, Sensitivity, Specificity

Procedure:

1: The original dataset $S$ is considered as the original layer of the deep sample space, and based on the original dataset, for each subject, the sample is pruned and the dataset $\hat{S}_p$ is obtained, and it is denoted as $P^0$;

2: FOR i =1: L DO

3:   FOR j=1: n DO

4: Sample pruning based on the envelope $p_j^{i-1}$ constructed from all sample segments of a single subject in

$$P^{i-1} = \begin{bmatrix} p_1^{i-1} \\ p_2^{i-1} \\ ... \\ p_n^{i-1} \end{bmatrix} \text{ to obtain } \hat{p}_j^{i-1};$$

5: Sample clustering using MlFC & IDMD algorithm to obtain a new prototype sample set $P_j^i$ the number of clusters is $C_i$;

6:   END

7: The set of $P^i$ is taken as the i-th layer of the deep sample space $P^i = \begin{bmatrix} p_1^i \\ p_2^i \\ ... \\ p_n^i \end{bmatrix}$;

8: END

9: Based on each layer $P^i$ of the deep sample space, the sample features are selected simultaneously for the constructed set of subject envelopes to obtain the $P_{ss}^i$;

10: The training dataset $P_{sstrain}^i$ and test dataset $P_{sstest}^i$ are divided by LOSO CV;

11: Train the SVM model using the training set and get the multilayer prediction labels $E$;

12: Sparse decision fusion multilayer prediction labels to get final labels $\hat{y}$;

13: Accuracy, sensitivity and specificity can be calculated;

## 3   Experimental results and analysis

### 3.1 Experimental conditions

#### 3.1.1 Datasets

In this paper, Sakar (dataset 1) [18], MaxLittle (dataset 2) [19] and the Self-harvested PD speech dataset (dataset 3) [51] are used as the target datasets for the validation of the algorithm. For the convenience of description, we refer to them as dataset 1, dataset 2, and dataset 3, respectively. These three datasets are highly representative, dataset 1 and dataset 2 are mainly used for diagnosis, and dataset 3 is mainly used for efficacy assessment. Dataset 1 and dataset 2 are public PD speech datasets based on European and American humans, and dataset 3 is a self-harvested PD speech dataset based on Chinese humans. Thus, these three datasets cover a representative range of humans from different applications and different regions. Table 1 shows the information of the three datasets.

Dataset 1 is the PD speech public dataset, which contains a total of 40 subjects, including 20 Parkinson's disease patients (6 females and 14 males) and 20 healthy subjects (10 females and 10 males). 26 speech sample segments were provided by each of the 40 subjects, and each speech sample segment contained 26 pronunciation tasks, specifically continuous vowels, digits, words, and phrases pronunciation, i.e., 26 features were extracted from each speech sample segment as a feature vector.

Dataset 2 was created by MaxLittle et al. This dataset contained a total of 31 subjects, 23 of whom had PD (7 females and 16 males) and 8 healthy (5 females and 3 males). Each subject contained 6 or 7 voice sample segments. Each speech sample segment contains 22 features and a feature vector is formed from these features.

Dataset 3 was provided by the First Affiliated Hospital of the Army Medical University, which contains a total of 90 subjects, including 36 PD patients (16 females and 20 males) and 54 healthy subjects (27 females and 27 males). Each subject was provided with 13 speech sample segments, each segment containing 26 speech features and forming a feature vector from these features.

Table 1. Datasets information

| Dataset | Number of subjects | Patients | Healthy people | Number of samples per subject | Samples | Futures | References |
|---|---|---|---|---|---|---|---|
| Sakar | 40 | 20 | 20 | 26 | 1040 | 26 | [18] |
| MaxLittle | 31 | 23 | 8 | 6 or 7 | 189 | 22 | [19] |
| SelfData | 90 | 36 | 54 | 13 | 1170 | 26 | [51] |

*3.1.2 Experimental conditions*

To ensure that a deep sample space can be constructed, the number of sample pruning is set based on the number of samples contained in each subject in the dataset. The deep sample space is generated by clustering each subject's sample envelope based on the previous layer, and the number of clusters is the number of samples contained in the previous layer's sample envelope minus one, and the resulting set of new prototype sample envelopes of subjects is used as the new sample space dataset.

The SVM, KNN and ELM classifiers were used in the experiments. The SVM used linear kernel function, the KNN took K=3, and the number of hidden neurons of ELM was set to 50.

In order to reflect the accuracy of the algorithm more comprehensively, two cross-validation methods, Leave One Subject Out (LOSO) and Holdout, were used for experimental validation, LOSO ensures that the training set and test set belong to different patients, and the classification accuracy is closer to the actual clinical diagnosis.

Equipment hardware and software environment: CPU (Intel i3-7100@3.90GHz), memory 8 GB, MATLAB R2018b.

*3.1.3 Evaluation Guidelines*

The evaluation criteria contain Accuracy (ACC), Sensitivity (SEN), Specificity (SPE) and Matthews Correlation Coefficient (MCC) [31].

$$Accuracy(ACC) = \frac{TP+TN}{TP+FP+TN+FN}$$

$$Sensitivity(SEN) = \frac{TP}{TP+FN}$$

$$Specificity(SPE) = \frac{TN}{FP+TN}$$

$$Matthews(MCC) = \frac{TP \times TN - FP \times FN}{\sqrt{(TP+FP)(TP+FN)(TN+FP)(TN+FN)}}$$

Where TP denotes true positive, FP denotes false positive, TN denotes true negative and FN denotes false negative.

## 3.2 Verification

### 3.2.1 Verification of the MlFC & SP_IDMD

Tables 7-8 show the validation of the algorithms in this paper on the generic dataset and the PD speech dataset. The "SVM" in the table denotes the classification of the dataset directly using the SVM classifier.

(1) Generic dataset

Table 2. Validity of MlFC & SP_IDMD in generic datasets

| DATASET | Method | ACC(%) | SEN(%) | SPE(%) |
|---|---|---|---|---|
| Yeast-1-2-8-9-vs-7 | SVM | 73.75 | 67.78 | 73.95 |
|  | Proposed algorithm +SVM | **74.56** | **71.11** | **74.67** |
| Yeast-0-5-6-7-9-vs-4 | SVM | 83.62 | 73.57 | 84.72 |
|  | Proposed algorithm +SVM | **86.63** | **75.00** | **87.92** |
| Wine | SVM | 94.73 | 91.36 | 96.97 |
|  | Proposed algorithm +SVM | **95.64** | **93.18** | **97.27** |
| WDBC | SVM | 94.83 | 90.47 | 97.41 |
|  | Proposed algorithm +SVM | **95.29** | **91.56** | **97.50** |
| Glass6 | SVM | 95.08 | 91.11 | 95.70 |
|  | Proposed algorithm +SVM | **96.31** | **91.11** | **97.14** |

(2) PD speech dataset

Table 3. Validity of MlFC & SP_IDMD in PD speech dataset (LOSO)

| Dataset | Methods | ACC(%) | SEN(%) | SPE(%) |
|---|---|---|---|---|
| Sakar | SVM | 53.37 | 52.31 | 54.42 |
|  | Proposed algorithm +SVM | **90.00** | **95.00** | **85.00** |
| MaxLittle | SVM | 66.67 | 75.00 | 58.33 |
|  | Proposed algorithm +SVM | **93.75** | **87.50** | **100.00** |
| SelfData | SVM | 51.82 | 47.01 | 56.62 |
|  | Proposed algorithm +SVM | **87.50** | **88.89** | **86.11** |

Table 4. Validity of MlFC & SP_IDMD in PD speech dataset (Holdout)

| Dataset | Methods | ACC(%) | SEN(%) | SPE(%) |
|---|---|---|---|---|
| Sakar | SVM | 64.81 | 64.10 | 65.51 |
|  | Proposed algorithm +SVM | **94.17** | **96.67** | **91.67** |
| MaxLittle | SVM | 73.53 | 62.86 | 84.29 |
|  | Proposed algorithm +SVM | **97.50** | **100.00** | **95.00** |
| SelfData | SVM | 56.14 | 56.36 | 55.93 |
|  | Proposed algorithm +SVM | **91.43** | **93.00** | **90.00** |

As shown in Tables 2-4, the algorithm proposed in this paper has a certain enhancement effect on each evaluation index in both the generic and PD speech datasets. Therefore, the algorithm not only can significantly improve the accuracy of PD speech diagnosis, but also has certain universality.

### 3.2.2 Verification of the deep sample learning mechanism

Table 5 shows that the deep sample learning mechanism has some error correction capability, as shown by subjects 23, 31, and 39, where the original layer prediction labels are inconsistent with the actual labels, but more than 80% of the layers from layer 2 to layer 6 are correctly predicted and can be finally corrected by the sparse fusion mechanism. However, there are also errors, as shown in subject 8, where the original layer prediction labels are consistent with the actual labels, but the final prediction labels obtained by fusing the multilayer prediction labels using the sparse fusion mechanism are inconsistent with the actual labels, resulting in misclassification. Overall, the number of successful error corrections is much greater than the number of misclassifications. Therefore, it can be seen that the deep sample learning mechanism has certain error correction capability and can improve certain recognition accuracy.

Table 5. Error correction performance of deep sample learning mechanism (Sakar)

| Subject num | Actual label | Predict label | Original Layer | Layer2 | Layer3 | Layer4 | Layer5 | Layer6 |
|---|---|---|---|---|---|---|---|---|
| 1 | 0 | 0 | 0 | 0 | 0 | 0 | 0 | 0 |
| 2 | 0 | 0 | 0 | 0 | 0 | 0 | 0 | 0 |
| 3 | 0 | 1 | 1 | 1 | 0 | 0 | 1 | 1 |
| 4 | 0 | 0 | 0 | 0 | 1 | 0 | 0 | 1 |
| 5 | 0 | 0 | 0 | 0 | 0 | 0 | 0 | 0 |
| 6 | 0 | 0 | 0 | 0 | 0 | 0 | 0 | 0 |
| 7 | 0 | 0 | 0 | 0 | 0 | 0 | 0 | 0 |
| **8** | **0** | **1** | **0** | **0** | **1** | **0** | **0** | **1** |
| 9 | 0 | 0 | 0 | 0 | 1 | 0 | 0 | 0 |
| 10 | 0 | 1 | 1 | 1 | 1 | 1 | 1 | 1 |
| 11 | 0 | 0 | 0 | 0 | 0 | 0 | 0 | 0 |
| 12 | 0 | 0 | 0 | 0 | 0 | 0 | 0 | 0 |
| 13 | 0 | 0 | 0 | 0 | 1 | 0 | 0 | 0 |
| 14 | 0 | 0 | 0 | 0 | 0 | 0 | 0 | 0 |
| 15 | 0 | 0 | 0 | 0 | 0 | 0 | 0 | 0 |
| 16 | 0 | 0 | 0 | 1 | 0 | 1 | 1 | 0 |
| 17 | 0 | 0 | 0 | 0 | 0 | 0 | 0 | 0 |
| 18 | 0 | 0 | 0 | 0 | 0 | 1 | 1 | 0 |
| 19 | 0 | 0 | 0 | 0 | 0 | 0 | 0 | 0 |
| 20 | 0 | 0 | 0 | 1 | 0 | 0 | 1 | 0 |
| 21 | 1 | 1 | 1 | 1 | 1 | 1 | 1 | 1 |
| 22 | 1 | 1 | 1 | 1 | 1 | 0 | 1 | 1 |
| **23** | **1** | **1** | **0** | **1** | **1** | **0** | **1** | **1** |
| 24 | 1 | 1 | 1 | 1 | 1 | 0 | 0 | 1 |
| 25 | 1 | 1 | 1 | 1 | 1 | 1 | 1 | 1 |
| 26 | 1 | 1 | 1 | 1 | 1 | 1 | 1 | 1 |
| 27 | 1 | 1 | 1 | 0 | 0 | 1 | 0 | 1 |
| 28 | 1 | 1 | 1 | 1 | 0 | 1 | 0 | 1 |
| 29 | 1 | 1 | 1 | 1 | 0 | 0 | 0 | 0 |
| 30 | 1 | 1 | 1 | 1 | 1 | 1 | 1 | 1 |
| **31** | **1** | **1** | **0** | **1** | **1** | **1** | **1** | **1** |
| 32 | 1 | 1 | 1 | 0 | 1 | 1 | 1 | 1 |
| 33 | 1 | 1 | 1 | 1 | 1 | 1 | 1 | 1 |
| 34 | 1 | 1 | 1 | 1 | 1 | 0 | 1 | 1 |
| 35 | 1 | 1 | 1 | 0 | 0 | 0 | 1 | 1 |
| 36 | 1 | 1 | 1 | 1 | 1 | 1 | 1 | 1 |
| 37 | 1 | 1 | 1 | 1 | 1 | 1 | 1 | 1 |
| 38 | 1 | 1 | 1 | 1 | 1 | 1 | 0 | 1 |
| **39** | **1** | **1** | **0** | **1** | **1** | **1** | **1** | **1** |
| 40 | 1 | 0 | 0 | 1 | 0 | 0 | 0 | 0 |

Note: red: successful error correction examples; yellow: misclassification examples

### 3.3 Algorithm Comparison

#### 3.3.1 Comparison with PD speech transformation algorithm

Sakar [18] used a simple linear sample transformation method that utilized concentrated trend and discrete measures such as mean, median, trimmed mean (removing 10% and 25%), standard deviation, interquartile range, and mean absolute deviation for 26 speech samples from each subject. The algorithm proposed in this paper is to prune the sample envelope of each subject and perform clustering based on the pruned sample envelope of the subjects, thereby generating a deep sample space. The sample features are selected simultaneously for each layer of the dataset in this space, thus achieving sample reconstruction to obtain a small number of samples of high quality. The experimental results of the comparison of the above two PD speech sample segment transformation algorithms are shown in Table 6.

Table 6. Comparison with PD speech sample segment transformation algorithm (Sakar)

| METHOD | ACC(%) | SEN(%) | SPE(%) |
|---|---|---|---|
| Sakar et al. [18] +KNN(3) +LOSO | 54.04 | 53.27 | 54.81 |
| Proposed algorithm +KNN(3) +LOSO | **90.00** | **90.00** | **90.00** |

| | | | | |
|---|---|---|---|---|
| Sakar et al. [18] +SVM +LOSO | | 52.50 | 52.50 | 52.50 |
| Proposed algorithm +SVM +LOSO | | **90.00** | **95.00** | **85.00** |

As shown in Table 6, the ensemble learning algorithm of enveloped deep FCM clustering of speech samples within subjects proposed in this paper is effective and has a large improvement in accuracy compared to the simple linear sample transformation.

### 3.3.2 Comparison with literature methods

Based on the Sakar, MaxLittle, and SelfData datasets, the results of this paper's algorithm compared with existing algorithms in the representative literature on PD diagnosis are shown in Tables 7-9.

Table 7. Comparison with previous PD diagnosis algorithms (Sakar)

| Study | Method | ACC (%) | SEN (%) | SPE (%) |
|---|---|---|---|---|
| Sakar et al. [18] | KNN +SVM | 55.00 (LOSO) | 60.00 | 50.00 |
| Canturk and Karabiber [32] | 4 Feature Selection Methods + 6 Classifiers | 57.50 (10-fold CV) | 54.28 | 80.00 |
| Zhang et al. [33] | MENN +RF with MENN | 81.50 (LOSO) | 92.50 | 70.50 |
| Benba et al. [35] | HFCC +SVM | 87.50 (LOSO) | 90.00 | 85.00 |
| Li et al. [35] | Hybrid feature learning +SVM | 82.50 (LOSO) | 85.00 | 80.00 |
| Benba et al. [36] | MFCC +SVM | 82.50 (LOSO) | 80.00 | 85.00 |
| Behroozi and Sami [37] | Multiple classifier framework | 87.50 (LOSO) | 90.00 | 85.00 |
| Zhang [38] | LSVM +MSVM +RSVM + CART +KNN +LDA +NB | 94.17 (Holdout) | 50.00 | 94.92 |
| Khan et al. [39] | Evolutionary Neural Network Ensembles | 90.00 (10-fold CV) | 93.00 | 97.00 |
| Zayrit et al. [40] | GA +SVM | 91.18 (10-fold CV) | — | — |
| Proposed algorithm | MlFC & SP_IDMD +SVM | **90.00 (LOSO)**<br>**94.17 (Holdout)**<br>**100.00 (10-fold CV)** | **95.00**<br>**96.67**<br>**100.00** | **85.00**<br>**91.67**<br>**100.00** |

Table 8. Comparison with previous PD diagnosis algorithms (MaxLittle)

| Study | Method | ACC (%) | SEN (%) | SPE (%) |
|---|---|---|---|---|
| Luukka. [41] | Fuzzy entropy measures +similarity | 85.03 (Holdout) | — | — |
| Spadoto et al. [42] | PSO +OPF harmony search +OPF gravitational search + OPF | 84.01 (Holdout) | — | — |
| Das [43] | ANN decision tree | 92.90 (Holdout) | — | — |
| Daliri [44] | SVM with Chi-square distance | 91.20 (Holdout) | 91.71 | 89.92 |
| Ali et al. [45] | DBN | 91.25 (Holdout) | 90.50 | 93.00 |
| Kadam and Jadhav [46] | FESA-DNN | 93.84 (Holdout) | — | — |
| Proposed algorithm | MlFC & SP_IDMD +SVM | **97.50 (Holdout)** | **100.00** | **95.00** |

Table 9. Comparison with previous PD diagnosis algorithms (SelfData)

| Study | Method | ACC (%) | SEN (%) | SPE (%) |
|---|---|---|---|---|
| Yang et al. [47] | KPCA-SVM | 59.67 (LOSO) | — | — |
| Galaz et al. [48] | SFFS-RF | 60.00 (LOSO) | — | — |
| Cigdem et al. [49] | ReliefF-FC-SVM | 62.67 (LOSO) | — | — |
| Spadoto et al. [50] | LDA-NN-GA | 63.00 (LOSO) | — | — |
| Liu et al. [51] | LPP +RF | 56.33 (LOSO) | 45.58 | 29.17 |
| Proposed algorithm | MlFC & SP_IDMD +SVM | **87.50 (LOSO)** | **88.89** | **86.11** |

For comparison, Table 7 presents the results of this paper's algorithm based on three cross-validation tests of LOSO, Holdout, and 10-fold CV. Table 8 presents the results of the algorithm in this paper based on the MaxLittle dataset using Holdout cross-validation tests. Table 9 presents the results of the algorithm in this paper based on the SelfData dataset using LOSO cross-validation tests. It can be seen that the algorithm in this paper still has some improvement in the evaluation indexes such as accuracy rate compared with the existing PD diagnosis representative literature algorithms, which proves the effectiveness of the algorithm in this paper.

### 3.4 Clinical analysis of the extracted features

In order to verify the contribution of this proposed algorithm to the clinical problem mentioned above, experiments on feature extraction is conducted. Figures 2-4 show the results obtained after sample reconstruction and simultaneous selection of sample features based on the three datasets Sakar, MaxLittle, and SelfData using the algorithm of this paper. Figure (a) shows a small number of samples obtained after reconstruction for each subject in this dataset, and the distribution of weights of different features in this sample segment is characterized by different color blocks. Figure (b) shows a uniform diagnostic marker that reflects the pathology of the whole subject, extracted from the weight distribution of the features in the different sample segments according to figure (a). Apparently, the number of the segments is smaller than that of the original segments, the complexity of the classification model is decreased largely. So, the generalization capability of the classification model is improved.

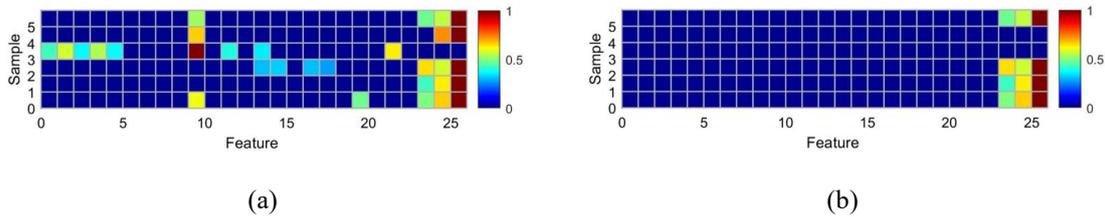

(a) (b)

Fig. 2. Weight map of features (Sakar):(a) weight map of features; (b)feature extraction map;

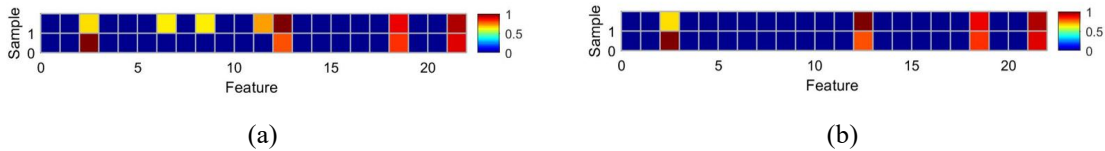

(a) (b)

Fig. 3. Weight map of features (MaxLittle):(a) weight map of features; (b)feature extraction map;

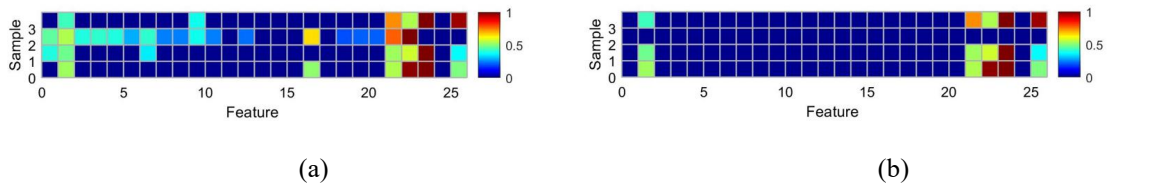

(a) (b)

Fig. 4. Weight map of features (SelfData): (a) weight map of features; (b) feature extraction map;

As can be seen from figure 2-4 (a) of the above three figures, the number of subjects' samples is significantly reduced and each sample segment has different salient features. Based on the Sakar dataset, after sample reconstruction using the algorithm of this paper, six high-quality samples were obtained for each subject. And as can be seen from Figure (b), for samples 1, 2, 3, and 6, features 24-26 can effectively characterize the pathology of the entire subject and can therefore be referred to as potential diagnostic markers. Similarly, in the MaxLittle dataset, 2 high-quality samples were obtained for each subject after sample reconstruction, and for these two samples, features 3, 13, 19, and 22 can be used as potential diagnostic markers. In the SelfData dataset, four high-quality samples were obtained for each subject after sample reconstruction, and for samples 1, 2, and 4, features 2, 22, 23, 24, and 26 can be used as potential diagnostic markers.

In summary, the algorithm in this paper allows sample reconstruction of the sample envelope of a subject to obtain a small number of samples of high quality. Based on these samples for simultaneous selection of sample features, uniform diagnostic markers that can characterize the pathology of this subject with high quality can be obtained. The algorithm not only helps to meet clinical needs but also improves diagnostic efficiency.

# 4 CONCLUSIONS

Parkinson's speech recognition is a non-invasive and efficient diagnostic modality that has become a research hotspot in recent years. The current existing PD diagnosis approaches are mainly based on the target dataset itself for improvement research, and do not consider the hierarchical structure information between sample ends. The current PD speech diagnosis also faces the following clinical application requirements:

1) It is difficult to reflect the Parkinson's pathological state comprehensively with a single corpus, and ensure the accurate diagnosis of PD speech since people have no enough priori knowledge. A large number of corpus samples (sample segments) need to be collected for each subject during PD speech diagnosis. However, when the number of samples collected for each subject is large, it will lead to a high complexity of the classification model and poor generalization ability. Therefore, it is necessary for sample reconstruction to reduce the number of segments collected for each subject, and improve the generalization ability of the classification model and the efficiency of PD speech diagnosis.

2) In the actual PD speech diagnosis, clinicians are interested in finding diagnostic speech markers that reflect the pathology of the whole subject. However, there are various segments within one subject, and the optimal relevant features of each speech sample segment are different to some extent. It is difficult for clinicians to find uniform diagnostic markers that reflect the pathology of the subjects. Therefore, it is necessary to combine the large segments into few segments, thereby finding the uniform optimal features (diagnostic markers).

Therefore, to solve the above mentioned problems, this paper proposes an ensemble learning algorithm of enveloped deep FCM clustering of speech samples within subjects of Parkinson's disease. Several representative PD speech datasets are selected and compared with the mainstream machine learning algorithms and the state-of-the-art related literature methods, respectively. The comparison results show that the reconstructed prototype segments of this paper's algorithm have higher recognition accuracy and stronger generalization performance. Besides, it helps to extract the speech features (diagnostic markers) reflecting the overall pathological status of the subject, which is more suitable for clinical applications.

The main contributions and innovations of the work in this paper are.

1) An enveloped deep sample learning network is proposed based on MlFCM and MIDMD-MlFC&IDMD: The MlFCM model is designed, and the interlayer consistency maintenance mechanism is constructed through the MIDMD. Combining the two parts above, an enveloped deep sample learning network is designed, thus obtaining a deep sample space characterizing the original samples, and has obtained a small number of high quality new samples.

2) The ensemble algorithm of PD speech data is designed based on MlFC&IDMD and SP module (called MlFC & SP_IDMD) is proposed: The sample set was pruned and then constructed as an envelope with all sample segments of a single subject. The envelope was processed using the deep sample learning network MlFC & IDMD to construct a PD speech sample segment deep space. The Sample segments and features in the envelop after MIFC&IDMD are simultaneously selected and sub-classifiers are constructed based on the selected segments and features. Finally, decision layer fusion is performed on all the sub-classifiers.

3) Based on the algorithm proposed in this paper, not only the classification accuracy and generalization performance can be improved, but also the features (diagnostic markers) reflecting the overall pathological status of the subject can be extracted. This helps physicians understand the mechanism of identifying PD by speech and also helps improve clinical applications. In other words, this proposed algorithm solves the clinical problems and needs mentioned above to some extent.

Although this paper demonstrates the effectiveness of the proposed method, further research and improvement are still needed. The algorithm improves evaluation metrics such as accuracy when the number of layers in the deep sample space is high, but it also enhances model complexity and computational time cost. Therefore, further optimization of the proposed algorithm can be considered in the future.